\documentstyle[emulateapj,graphicx]{article}
\def\etal   {{et~al.}\ }
\def\msun{{\rm\,M_\odot}}
\def\Zsun{{\rm\,Z_\odot}}

\def\vol#1  {{{#1}{\rm,}\ }}
\def\lya{{\rm Ly}\alpha}

\def\etal{et al.\ }

\newcount\refno
\newcount\rfno
\def\eq{$^{\the\refno\ }$\advance\refno by 1}
\def\ad{\advance\rfno by 1}

\def\clock{\count0=\time \divide\count0 by 60
     \count1=\count0 \multiply\count1 by -60 \advance\count1 by \time
     \number\count0:\ifnum\count1<10{0\number\count1}\else\number\count1\fi}

\def\myputfigure#1#2#3#4#5%
{\hskip0.03\textwidth\vskip#5pt%\hfill
\makebox[0pt]{\hskip#2in
\includegraphics[width=#3\textwidth]{#1}}\vskip#4pt\hfill}
\newcommand{\beq}{\begin{equation}}
\newcommand{\eeq}{\end{equation}}


\begin{document}
\title{Revealing the Warm-Hot Intergalactic Medium with OVI Absorption}
\author{Renyue Cen, Todd M. Tripp, Jeremiah P. Ostriker, and Edward B. Jenkins}
\centerline{Princeton University Observatory, Princeton University, Princeton, NJ 08544}
\centerline{cen@astro.princeton.edu, 
tripp@astro.princeton.edu}
\centerline{jpo@astro.princeton.edu,
ebj@astro.princeton.edu}

\begin{abstract}

Hydrodynamic simulations of growth of cosmic structure 
suggest that 30-50\% of the total
baryons at $z=0$ may be in a warm-hot intergalactic medium (WHIM)
with temperatures $\sim 10^5-10^7$K.
The \ion{O}{6} $\lambda \lambda$1032, 1038 
 absorption line doublet 
in the FUV portion of QSO spectra
provides an important probe of this gas.
Utilizing recent hydrodynamic simulations,
it is found that
there should be $\sim 5$ O~VI absorption lines per unit redshift
with equivalent widths $\ge 35~{\rm m\AA}$,
decreasing rapidly to $\sim 0.5$ per unit redshift at $\ge 350~{\rm m\AA}$.
About 10\% of the total baryonic matter or $20-30$\% of the WHIM 
is expected to be in the O~VI absorption line systems
with equivalent width $\ge 20~{\rm m\AA}$;
the remaining WHIM gas may be too hot or have too low
metallicity to be detected in O~VI.
We find that the simulation results agree well
with observations with regard to the line abundance
and total mass contained in these systems.
Some of the O~VI systems are collisionally ionized and some are 
photoionized, but most of the mass is in the collisionally ionized systems.
We show that the gas that produces
the O~VI absorption lines does not reside in virialized
regions such as galaxies, groups, or clusters of galaxies,
but rather has an overdensity of $10-40$ times the average density.
These regions form a somewhat connected network of filaments.
The typical metallicity of these regions is $0.1-0.3\Zsun$.

\end{abstract}


\section{Introduction}

Three independent observations at $z\ge 2$,
namely 
the latest cosmic microwave background experiments 
(Netterfield \etal 2001) at $z>>2$,
the Lyman alpha forest decrement measurements
(Rauch \etal 1997; Weinberg \etal 1997) at $z\sim 2-3$,
and the deuterium-to-hydrogen ratio measurements
(interpreted within the context of the standard nucleosynthesis theory,
Burles \& Tytler 1998),
all consistently give 
$\Omega_{b} (z\ge 2) = (0.019\pm 0.002)h^{-2} = 0.039\pm 0.004$
(where $h=H_0/100~{\rm km~sec^{-1}~Mpc^{-1}}=0.7$ 
is used for the last equality).
On the other hand,
in our local universe all the well-measured baryonic components
do not appear to add up to the indicated baryonic density 
seen at high redshift 
by a factor of about three 
(e.g., Fukugita, Hogan, \& Peebles 1998)
with 
$\Omega_{b}(z=0) =\Omega_{*} + \Omega_{HI} + \Omega_{H_2} + \Omega_{Xray,cl} \approx 0.0068 \le 0.011~~(2\sigma~\hbox{limit})$,
where $\Omega_{*}$, $\Omega_{HI}$, $\Omega_{H_2}$ 
and $\Omega_{Xray,cl}$ are the baryonic densities
contained in stars, neutral atomic hydrogen detected in 21cm or damped
Lyman alpha systems, molecular hydrogen
and hot X-ray emitting gas in rich cluster centers.
While a substantial amount of baryons may be
in the low-redshift $\lya$ forest with $\Omega_{\lya} = 0.008$
(Penton, Shull, \& Stocke 2000), it is still far short of accounting for
the baryons seen at high redshift.
Thus, most of the baryons are ``missing" today.

  Hydrodynamic simulations of structure growth suggest that a 
  large fraction of the missing baryons at the present epoch may 
  be in a gaseous phase with temperatures of $\sim 10^{5} - 
  10^{7}$ K at moderate overdensities, typically $\sim 10 - 40$ 
  (Cen \etal 1995; 
  Cen \& Ostriker 1999a; Dav\'{e} et al. 2001). It is important to 
  detect this gas to enable a better understanding of the baryon 
  distribution in the local universe and because it may have an 
  important influence on the evolution of galaxies (Blanton et 
  al. 2000). The \ion{O}{6} $\lambda \lambda$1032, 1038 
  absorption line doublet in the spectra of low-redshift QSOs 
  provides a valuable probe of the quantity and properties of 
  this gas, and recent observations with the Space Telescope 
  Imaging Spectrograph on the {\it Hubble Space Telescope} have 
  shown that the number of intervening \ion{O}{6} absorbers per 
  unit redshift ($dN/dz$) is quite high (Tripp, Savage, \& 
  Jenkins 2000; Tripp \& Savage 2000),
  suggesting that these 
  absorbers do indeed trace a significant baryon repository. In 
  this {\it Letter} we use cosmological hydrodynamic simulations 
  to make detailed predictions of the \ion{O}{6} absorption line 
  properties, and we compare these predictions to recent 
  observations.
  This complements earlier work 
by Hellsten, Gnedin and Miralda-Escud\'e (1998),
which focused on higher ionization oxygen lines 
observable with X-ray telescopes.

\section{Simulation}

We use two recent cosmological hydrodynamic simulations
of the canonical, 
cosmological constant dominated
cold dark matter model (Ostriker \& Steinhardt 1995)
with the following parameters:
$\Omega_m=0.3$, $\Omega_{\Lambda}=0.7$, $\Omega_b h^2=0.017$, $h=0.67$, 
$\sigma_8=0.9$, and  the spectral index of the primordial 
mass power spectrum $n=1.0$. 
The higher resolution 
simulation box has a size of 25$h^{-1}$ comoving Mpc on 
a uniform mesh with $768^3$ cells and 
$384^3$ dark matter particles.
The comoving cell size is $32.6h^{-1}$ kpc. 
The mass of each dark matter particle
is $2.03\times 10^7\msun$, 
and the mean baryonic mass per cell is 
$3.35\times 10^5h^{-1}M_\odot$.
An earlier simulation with a coarser resolution but a larger 
box of size $L=100h^{-1}$Mpc (Cen \& Ostriker 1999a,b)
is also used to calibrate the boxsize effect.
In velocity units the cell sizes are
$\Delta v=(3.3,19.5)~$km/s in the two simulations, respectively,
providing adequate spectral resolution in comparison to
current observations.\footnote{The STIS observations reported by
Tripp \etal 2000 and Tripp \& Savage 2000 have
spectral resolution of $7$km/s (FWHM).}

A word on simulation resolution is relevant here.
As simulations with much higher spatial resolutions have shown
(Dave et.al. 2001), the bulk of the warm-hot intergalactic 
medium has an overdensity of $10-40$. 
This has an important implications: the WHIM gas is mostly in uncollapsed
regions, for which the Jeans mass is well resolved by our high resolution
simulation. 
In fact, even for a lower temperature, photoionized gas at 
$T=10^4~$K, the Jeans mass is $\sim 10^9\msun$ and Jeans length is 
$400~$kpc/h comoving, which are to be compared to the 
$3.3\times 10^5\msun$ and $2\times 10^7\msun$ 
nominal baryonic and dark matter mass resolution, and 
$33$kpc/h comoving nominal spatial resolution of our 
$25$Mpc/h box; the corresponding 
numbers for our $100$Mpc/h box are 
$7.1\times 10^7\msun$ and $5.1\times 10^9\msun$, and $195$kpc/h. 
The true spatial and mass resolutions of the simulation 
are probably a factor of 2 and 8 times worse (Cen \& Ostriker 1999b).
Clearly, our higher resolution simulation
is adequate for resolving unvirialized intergalactic gas that was 
initially photoionized and then shock heated to the indicated temperature.
Our lower resolution simulation is marginal at resolving
smaller WHIM structures, which will be evident in the
results shown below.

We follow star formation and feedback using a well defined prescription
similar to that of other investigators.
A stellar particle of mass
$m_{*}=c_{*} m_{\rm gas} \Delta t/t_{*}$ is created,
if the gas in a cell at any time meets
the following three conditions simultaneously:
(i) flow contracting, (ii) cooling time less than dynamic time, and 
(iii) Jeans unstable,
where $\Delta t$ is the timestep, $t_{*}={\rm max}(t_{\rm dyn}, 10^7$yrs),
$t_{dyn}$ is the dynamical time of the cell,
$m_{\rm gas}$ is the baryonic gas mass in the cell and
$c_*=0.25$ is star formation efficiency.
Each stellar particle has a number of other attributes at birth, including 
formation time and initial gas metallicity.
The typical mass of a stellar particle
is one million solar masses.
Stellar particles 
are subsequently treated dynamically
as collisionless particles,
except that feedback from star formation is allowed in
three forms: UV ionizing field, supernova kinetic energy, and metal enriched
gas, all being proportional to the local star formation rate.
Metals are ejected into the local gas cells where
stellar particles are located using a yield $Y=0.02$
(Arnett 1996) and are followed
as a separate species adopting the standard solar composition.
The simulation also includes cooling/heating processes due to 
the self-consistently produced metals using a code based
on the Raymond-Smith code
assuming ionization
equilibrium (Cen \etal 1995).
We find that the computed metallicity distributions
over a wide range of environments, including 
clusters of galaxies, damped Lyman systems, Lyman alpha forest
and stars, are in broad agreement with observations
(Cen \& Ostriker 1999c; Nagamine \etal 2001; Cen \etal 2001),
lending us confidence that the computed metal distribution
in the intermediate regions 
under consideration 
(between Lyman alpha forest and clusters of galaxies)
may be a good approximation to the real universe.

Post-simulation the photoionization code CLOUDY (Ferland et al. 1998) is
used to compute the abundance of O~VI,
adopting the shape of the UV
background calculated by Haardt \& Madau (1996) normalized by the
intensity at 1 Ryd determined by Shull et al. (1999) and assuming
ionization equilibrium.
We generate synthetic absorption spectra 
using a code similar to that used 
to generate the synthetic Lyman alpha forest
spectra in earlier papers
(Cen \etal 1994; Miralda-Escud\'e \etal 1996).
To elucidate the physical nature of the absorbers,
we distinguish between photoionized and collisionally ionized gas;
O~VI absorption lines with optical depth-weighted
temperature less than $10^5~$K are identified as photoionized
systems and others as collisionally ionized lines.
However, 
in some cases both ionization mechanisms are important.

\vspace{1.0cm} 
\myputfigure{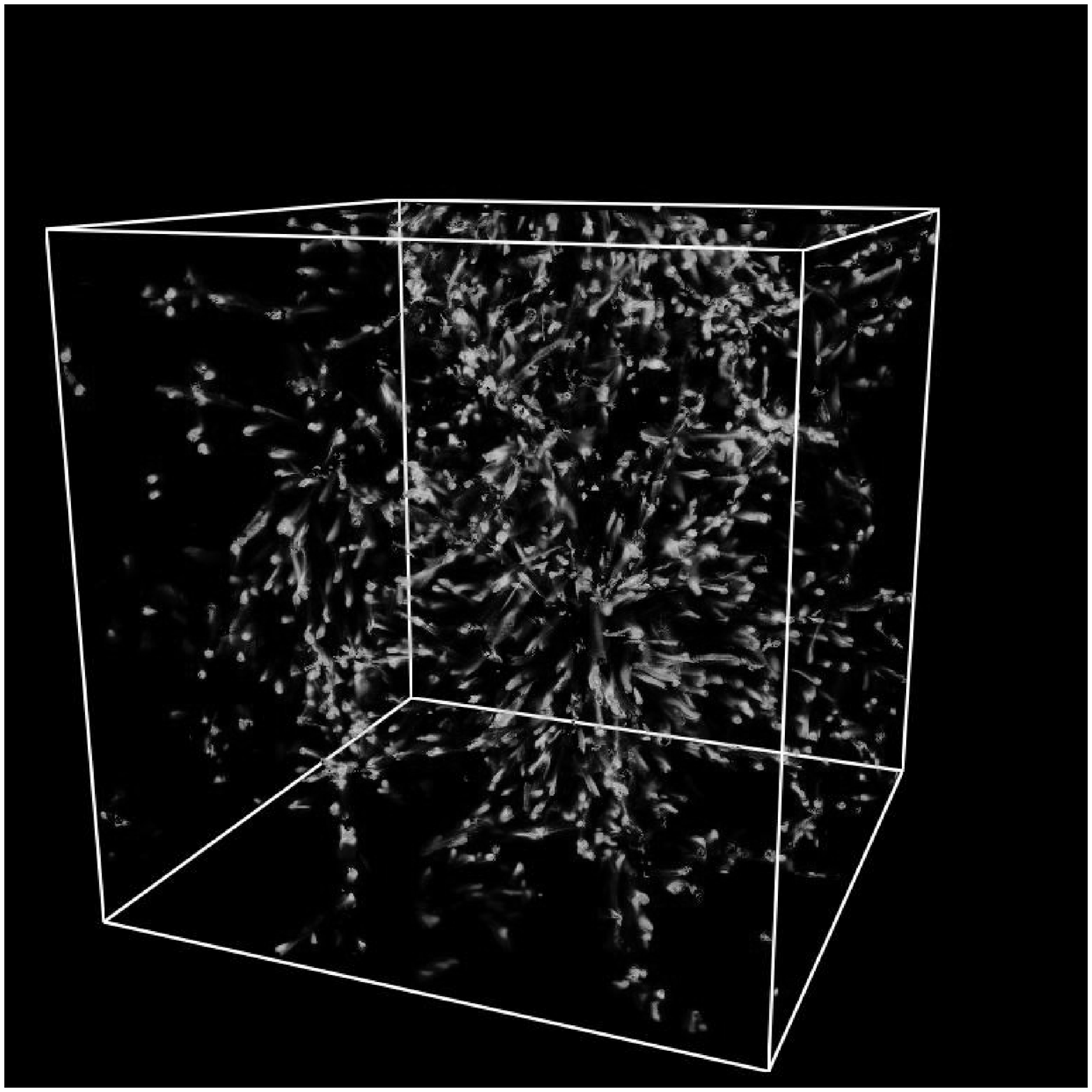}{3.2}{0.45}{-10}{-10} 
\vspace{1.0cm} 
\myputfigure{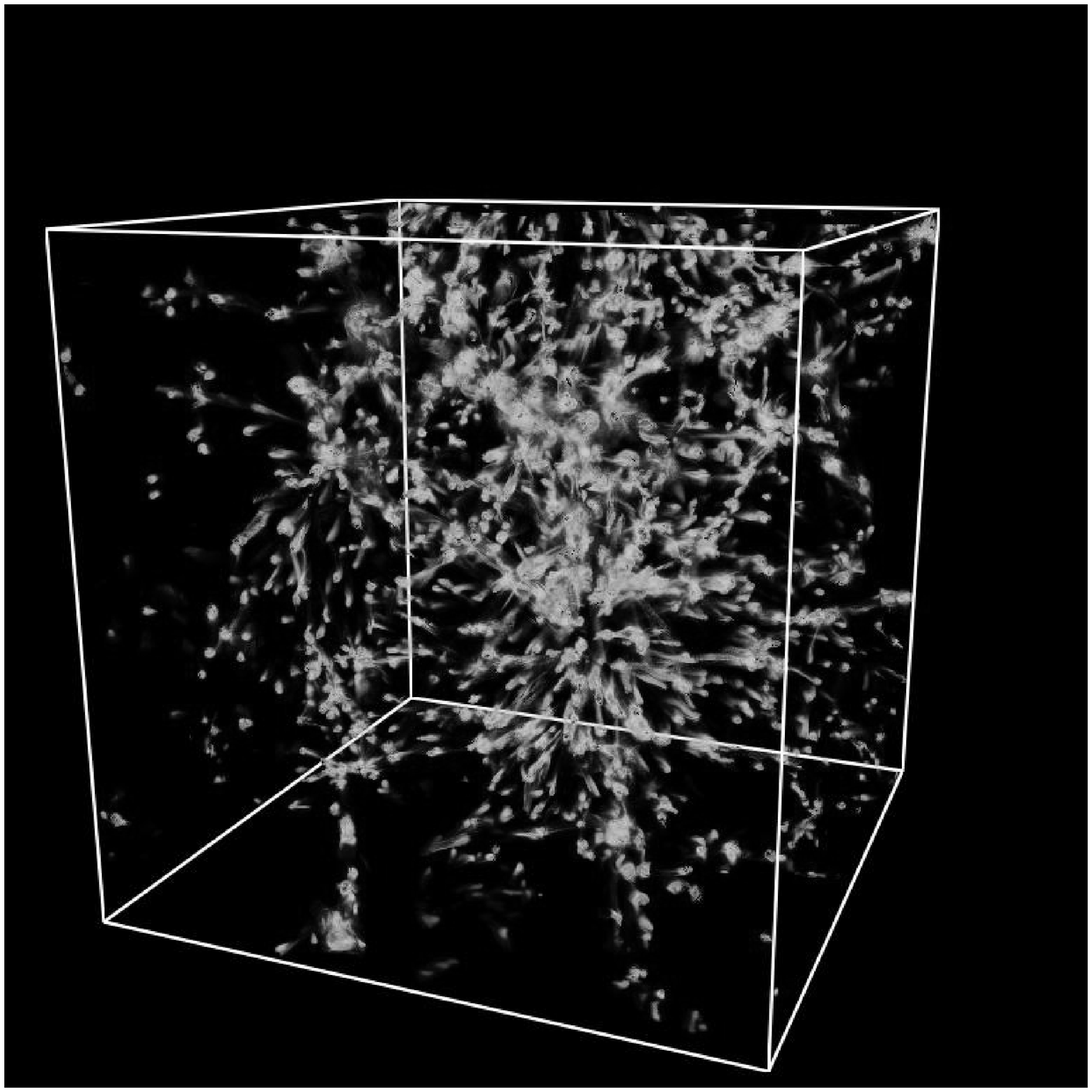}{3.2}{0.45}{-10}{-10} 
\vspace{1.2cm} 
\vspace{-0.5cm} 
\figcaption{
Three-dimensional distribution of the O~VI density 
for photoionized gas (top panel) and collisionally ionized gas (bottom panel).
The box size is $25h^{-1}$Mpc.
The color scheme is as follows:
light green,
green,
bright green,
green-yellow,
yellow,
yellow-red and
red-black regions
have O~VI densities of 
$10^{-7}$, 
$10^{-6.5}$, 
$10^{-6}$,
$10^{-5.5}$, 
$10^{-5}$, 
$10^{-4.5}$ 
and $> 10^{-4.5}$ in units of $\rho_{crit}$.
\label{fig:tau}}
\vspace{\baselineskip}

\section{Results}

Figure 1 shows the 3-d distribution of the O~VI density:
Figure 1a shows the photoionized O~VI and 
Figure 1b shows the collisionally ionized O~VI.
It is seen that the low density (bright green) O~VI gas 
is somewhat connected to form a filamentary structure,
while higher density gas surrounds 
roundish regions. 
Galaxies usually sit in the middle of the roundish red regions 
(Cen \& Ostriker 1999c).
The O~VI lines due to photoionized gas tend to stem from 
lower density regions.
As we will show later, the bright-green and green-yellow
regions produce most of the observed
OVI absorption lines;
these regions have overdensities of $10-40$
times the average density.

\myputfigure{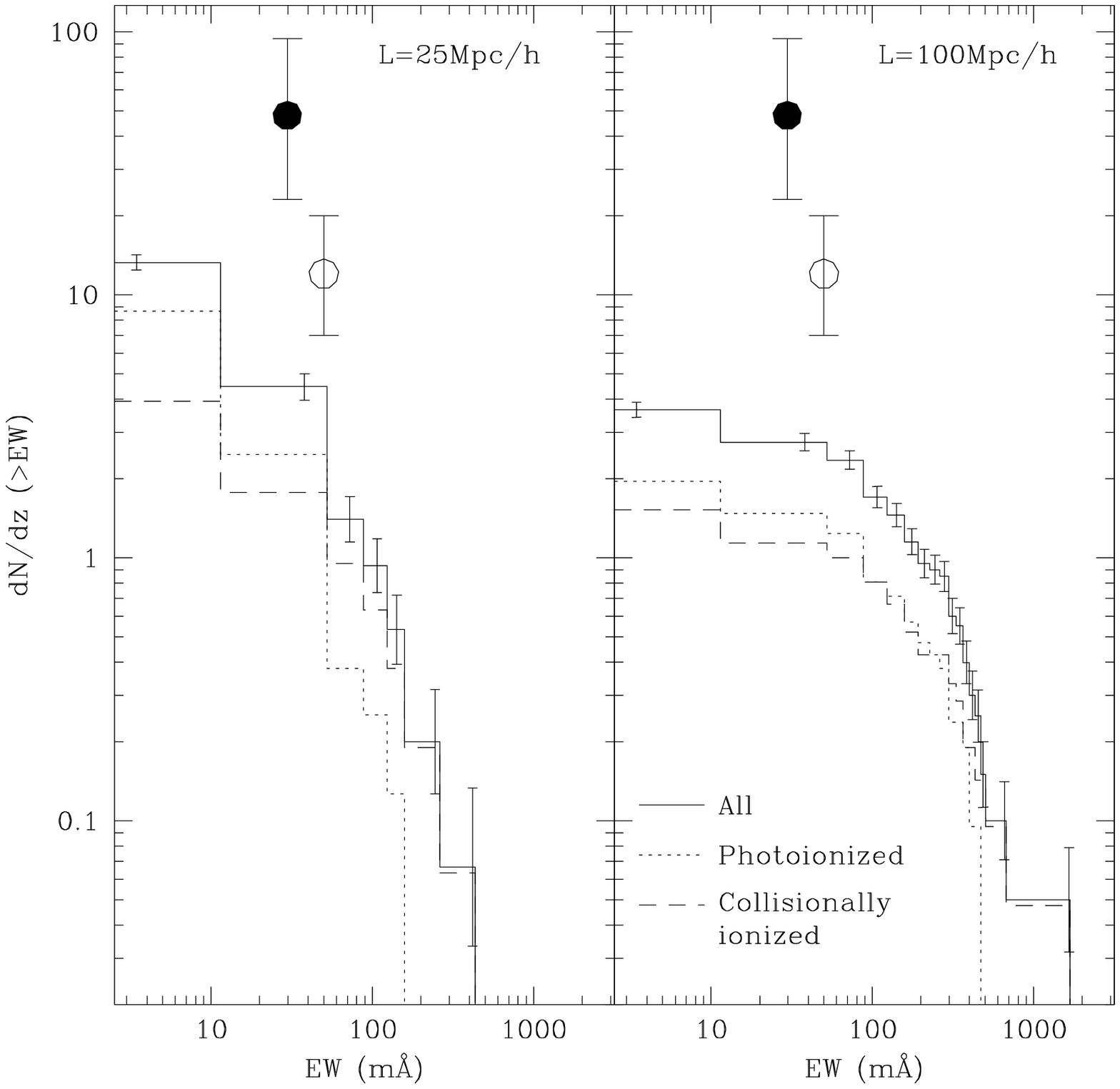}{3.2}{0.45}{-10}{-10} 
%\vspace{-3.5cm} 
\vspace{-0.5cm} 
\figcaption{
Cumulative number of OVI absorption lines
per unit redshift as a function of 
the O~VI absorption line
equivalent width
for all lines (solid curve), lines due to photoionized gas (dotted curve)
and 
lines due to collisionally ionized gas (dashed curve) with
$1\sigma$ statistical errorbars.
The left panel is from the $L=25h^{-1}$Mpc box 
and the right panel is from the $L=100h^{-1}$Mpc box.
The equivalent widths are computed by integrating the flux spectrum 
between downcrossing and upcrossing points at 80\% of the continuum.
Both dotted and dashed curves are shifted downward by 5\% 
for clarity.
The symbols are observational data from 
Tripp, Savage, \& Jenkins (2000; solid dot)
and Savage \etal (2001; open dot) with $1\sigma$ errorbars.
\label{fig:tau}}
\vspace{\baselineskip}

Figure 2 shows cumulative number of lines per unit redshift
as a function of the O~VI absorption line
equivalent width for all lines (solid curve),
photoionized lines (dotted curve) and collisionally 
ionized lines (dashed curve),
compared to observations from Tripp \etal (2000)
and Savage \etal (2001).
We note that at $EW<35{\rm m\AA}$ the photoionized lines
outnumber the collisionally ionized lines.
But at the high end of EW 
the collisionally ionized lines dominate in total numbers of lines,
due to the fact that high EW lines 
originate only in high temperature and higher density regions.
Visual examinations of synthetic O~VI absorption spectra indicate that 
O~VI absorption lines due to photo-ionized gas are typically  
narrower ($b$ parameter) than 
O~VI absorption lines due to collisionally ionized gas.
%that contain more gas and 
%also tend to have greater bulk motions.
%the collisionally ionized lines 
The agreement between simulations and observations is reasonably
good, given the small observational samples and computational
issues which may have caused both boxes to underestimate
the number of lines at various equivalent widths.
The larger simulation box (right panel) 
definitely underestimates the number of lines
at $EW\le 70{\rm m\AA}$ due to inadequate resolution,
while even the smaller box may be underestimating  
the number of lines at $EW\le 20{\rm m\AA}$ 
and still higher spatial resolution (in progress)
would be required to adequately model these lines.
But, at larger width, cosmic variance is substantial
and the smaller box underestimates the abundance of 
lines with $EW\ge 350{\rm m\AA}$.
We argue that a box of size $L>100$Mpc/h may be required,
if one is interested in studying lines with $EW\ge 350{\rm m\AA}$.
We expect that improvement of numerical simulations
will likely increase the computed number of lines
at both ends of the equivalent width distribution.
Recently, Fang \& Bryan (2001) have independently obtained
very similar results from a cosmological simulation
which employs different numerical methods.

\myputfigure{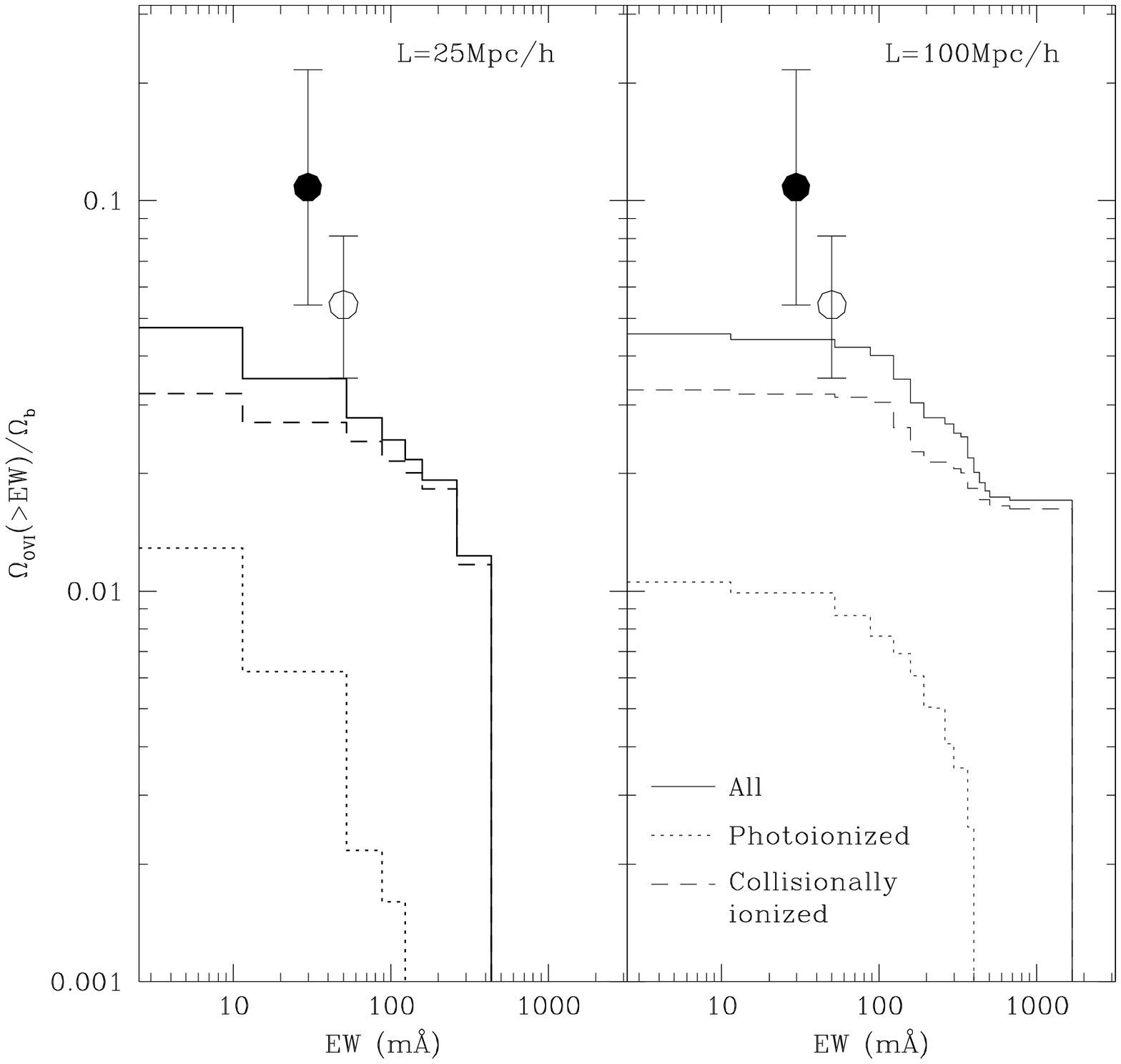}{3.2}{0.45}{-10}{-10} 
%\vspace{-3.5cm} 
\vspace{-0.5cm} 
\figcaption{
Cumulative distributions
of the fraction of baryonic gas contained 
in OVI absorption line, again, for all lines
(solid curve), lines due to photoionized gas (dotted curve)
and 
lines due to collisionally ionized gas (dashed curve).
The left panel is from the $L=25h^{-1}$Mpc box 
and the right panel is from the $L=100h^{-1}$Mpc box.
Both dotted and dashed curves are, again, shifted downward by 5\% 
for clarity.
The symbols are observational data from 
Tripp, Savage, \& Jenkins (2000; solid dot)
and Savage \etal (2001; open dot) with $1\sigma$ errorbars,
assuming $(Z/\Zsun) f(OVI)=0.02$ adopted from simulation (see the bottom panel
of Figure 4)
\label{fig:tau}}
\vspace{\baselineskip}

Figure 3 shows the cumulative fraction of baryonic gas density contained 
in the O~VI lines 
as a function of equivalent width for all lines (solid curve),
photoionized lines (dotted curve) and collisionally
ionized lines (dashed curve), compared
to observational constraints.
Apparently, most of the mass probed by O~VI lines
is in lines that are collisionally ionized,
although the photoionized lines are dominant in number.
This is because collisionally ionized lines originate 
in higher density regions (see Figure 4 below).
We see that the baryonic gas contained in O~VI lines
is about 10\% of total baryonic matter or roughly $20-30$\% of the WHIM.
The agreement with observations 
is good, together with the agreement
found in Figure 2 strongly suggesting 
the existence of the WHIM in the real universe
and that the inventory of mass derived from the
O~VI absorption lines accounts for a significant fraction of WHIM.
The remaining 70-80\% of the WHIM is likely 
gas which it too hot or does not contain
enough oxygen to be detected via O~VI absorption.

Finally, Figure 4 reveals the physical properties 
of the underlying gas that produces the O~VI lines.
We see that at $EW=(35,100){\rm m\AA}$ the typical metallicity
of the gas that produces  O~VI absorption lines
is $0.1-0.3\Zsun$ for both photoionized lines
and collisionally ionized lines.
The typical overdensity is $\sim (10-40)$ for both photoionized 
collisionally ionized lines in the range $EW=20-200{\rm m\AA}$.
Photoionized lines tend to reside at the lower end of the overdensity range
at the lower EW end,
but two types converge 
to a typical overdensity of roughly $40$ 
at the high EW end under consideration.
Evidently, the gas that causes the O~VI absorption lines
is relatively distant from virialized regions such as galaxies,
groups, or clusters of galaxies
in truly intergalactic, uncollapsed regions.

\myputfigure{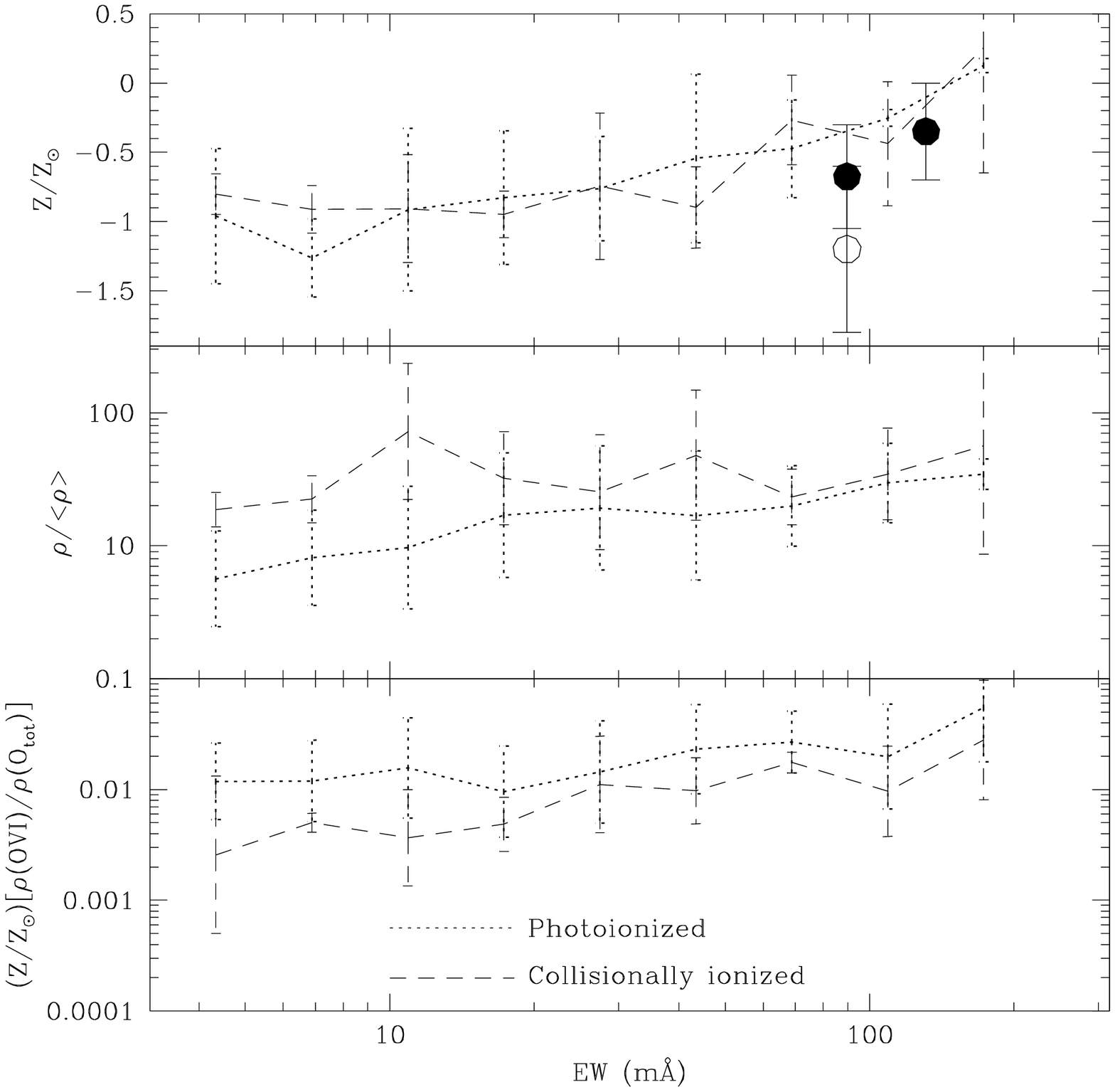}{3.2}{0.45}{-10}{-10} 
%\vspace{-3.5cm} 
\vspace{-0.5cm} 
\figcaption{
The physical properties 
of the underlying gas that produces the O~VI lines.
The top panel shows metallicity of O~VI lines
as a function of equivalent width 
for lines due to photoionized gas (dotted curve)
and lines collisionally ionized gas (dashed curve),
with $1\sigma$ dispersion.
The symbols are observed metallicities derived by 
Tripp \etal (2001) and Savage \etal (2001) assuming
the gas is photoionized (solid dots)
or collisionally ionized (open dot)
with $2\sigma$ errorbars.
The middle panel shows the overdensity of regions
that produce the O~VI absorption lines 
as a function of equivalent width 
for lines due to photoionized gas (dotted curve)
and due to collisionally ionized gas (dashed curve)
with $1\sigma$ dispersion.
The bottom panel shows the product of 
metallicity and O~VI fraction
as a function of equivalent width
for lines due to photoionized gas (dotted curve)
and lines collisionally ionized gas (dashed curve) 
with $1\sigma$ dispersion.
This quantity is useful for calculating $\Omega_b(OVI)$
from observations, see eqn. 1 of Tripp \etal (2000).
\label{fig:tau}}
\vspace{\baselineskip}

\section{Discussion and Conclusions}

We have made detailed analyses of our latest hydrodynamic simulations
to examine the expected properties of O~VI absorption lines due to 
intergalactic gas.
In the simulations 
the vast majority of O~VI absorption lines 
with equivalent width of $\ge 35{\rm m\AA}$
are not due to gas that resides in virialized
regions such as galaxies, groups, or clusters of galaxies,
but rather due to intergalactic gas of an overdensity of $10-40$.
These regions form a somewhat connected network of filaments.
The typical metallicity of these regions is
$0.1-0.3\Zsun$.
A relevant test for this contribution is to check
in both the real and simulated sky:
what is the distance between O~VI absorption systems and bright galaxies?

We find that the simulation results agree well
with observations with regard to the line abundance
and total mass contained in these systems.
At O~VI absorption line equivalent width of $\ge 35{\rm m\AA}$
one expects to see $\sim 5$ lines per unit redshift.
The number of lines decreases rapidly towards larger equivalent
width, dropping to about $0.5$ per unit redshift at $\ge 350{\rm m\AA}$,
above which our current simulations may be significantly 
underestimating the line abundance.
About 10\% of total baryonic matter or $20-30$\% of WHIM 
is expected to be in the O~VI absorption line systems
with equivalent width $\ge 20{\rm m\AA}$.
Clearly, a larger observational sample of O~VI absorption lines
would be highly desirable
to increase the statistical accuracy as well
as other independent observations including
absorption lines due to other species and direct emission 
measurements.

The evolution of WHIM is likely a function of cosmological models,
and to some extent, parallels the evolution 
of clusters of galaxies thus provides a powerful, independent test of 
cosmological models.
The model appears to be qualitatively
consistent with the latest observations at higher redshifts
(Reimers \etal 2001).
We will explore this issue of evolution of WHIM quantitatively
in a separate paper.

\acknowledgments
RC and JPO acknowledge support for this research 
by NASA grant NAG5-2759 and NSF grants AST93-18185, ASC97-40300 and
TMT and EBJ acknowledge support for this research from NASA through
grants GO-08165.01-97A and GO-08695.01-A from the Space Telescope Science
Institute.
We thank the referee for bringing up an important issue regarding
the numerical resolution, which helped us more clearly state
the relevant effects.

\end{document}